\documentstyle[12pt,aps]{revtex}

\begin{document}
\draft
\title{Vector-like Fermion Doubling and Pair-creation on a lattice}
\author{She-Sheng Xue
}
\address{
ICRA, INFN  and
Physics Department, University of Rome ``La Sapienza", 00185 Rome, Italy
}


\maketitle

\centerline{xue@icra.it}

\begin{abstract}

Based on the Pauli principle and Heisenberg uncertainty 
relationship, we show that the phenomenon of 
vectorlike fermion-doubling in quantum field theories on a lattice
is related to the phenomenon of pair-creation of massless fermions, 
due to quantum-field fluctuations upon the zero-energy level of 
the ground state of regularized quantum field theories on the lattice. 

\end{abstract}


\narrowtext

The standard model for particle physics is strongly phenomenologically supported 
in low-energy. Neutrino fields in this model are purely left-handed and their 
gauge couplings are entirely parity-violating. The renormalizable quantum 
field theory describing the standard model is thus a chiral gauge theory with 
parity-violating gauge symmetries and fermion content. 
However, as a ``no-go" theorem, the Nielsen and M.~Ninomiya \cite{nn81} 
demonstrated that this type of quantum field theories 
cannot be consistently regularized and quantized on the lattice 
for the phenomenon of vectorlike fermion-doubling, i.e., both left-handed and 
right-handed neutrinos inevitably appear in the low-energy spectrum of 
the theory. This has been a fundamental problem in particle physics for two decades. 
Broadly speaking, this ``no-go" theorem is demonstrated by 
using the continuity of dispersion relations in a Brillouin zone or 
cancelation of gauge anomalies. To have a further
physical insight on this ``no-go" theorem, we try to understand it in terms of 
fermionic nature of the ground state of regularized quantum field theories, 
based on the Pauli principle and Heisenberg uncertainty relationship.   

In this brief report, we point out that the phenomenon of 
vectorlike fermion-doubling is in fact the 
phenomenon of pair-creations of massless fermion and antifermion, 
due to quantum-field
fluctuations around the zero-energy level ($E=0$) of a 
regularized vacuum of quantum field theories for fermions. The regularized 
vacuum indicates the ground state of regularized quantum field theories for 
free fermions, whose degree of freedoms is {\it finite}.
Here, we adopt the lattice regularization. The phenomenon 
of pair-creation indicates that a pair of massless fermion and antifermion is 
created out of the regularized vacuum and exhibits as low-energy excitations
in the spectrum of the theory.  

As an example for illustrating 
this without losing any generality, we take a free field theory of a 
left-handed chiral fermion $\psi_L$ on the 1+1 dimensional spacetime lattice. The
vacuum of lattice-regularized quantum field theories is henceforth indicated as 
the lattice vacuum. With the fixed spatial and temporal lattice spacings 
$a$ and $a_t$ $(a\gg a_t)$, 
the free Hamiltonian of the left-handed chiral fermion $\psi_L$ is given by,
\begin{equation}
H_\circ ={1\over 2a}\sum_x\left(\bar\psi_L(x) \partial\cdot\gamma\psi_L(x)\right),
\label{efree}
\end{equation}
where the fermionic field is two-component and dimensionless Weyl
field, $x$ is the integer label of space sites, $\gamma$-matrix ($\gamma^2=1$) and
\begin{equation} 
\partial=(\delta_{x,x+1}
-\delta_{x,x-1}),\hskip0.3cm
\delta_{x,x\pm 1}\psi_L(x)=\psi_L(x\pm 1).
\label{ekinetic} 
\end{equation}
This chiral fermion has charges and appropriately couple to gauge fields. 
To simplify discussions, we define the physical momentum as $\tilde p$ and 
dimensionless momentum $p\equiv a\tilde p$.

This Hamiltonian for fermion field $\psi_L$ has negative-energy states. The 
lattice vacuum is defined by an ensemble of all fully occupied negative-energy 
states and zero-energy state of the free Hamiltonian (\ref{efree}) 
in the energy range of $0\ge E\ge-\pi/a_t$ ($\hbar=1$), as represented 
in the $E\le 0$ part of the energy-momentum ($E-p$) plane.
According to the Pauli principle, 
the full occupation of all negative-energy states and zero-energy state 
means that the zero-energy state and each negative-energy state, 
within $0\ge E\ge-\pi/a_t$, are filled by both particle $\psi_L$ and its 
antiparticle $\psi_L^c$ states with the opposite charges and chiralities,
\begin{equation} 
\psi_L:\hskip0.2cm E=-\tilde p,\hskip0.2cm \tilde p\ge 0; \hskip0.4cm 
\psi^c_L:\hskip0.2cm E=\tilde p,\hskip0.2cm \tilde p\le 0.
\label{elr} 
\end{equation}
Thus, the lattice vacuum exactly preserves chiral symmetries of the free
Hamiltonian (\ref{efree}). The volume of the lattice vacuum in the 
energy-momentum plane ($E-p$), i.e., $0\ge E\ge-\pi/a_t$ and $|a\tilde p|\le\pi$, 
is {\it finite}. 
As a consequence, the total number of negative-energy states must be {\it finite} for 
the reason that each state occupies a quantum volume of $h$ or $2\pi$ for 
$\hbar=1$ in the energy-momentum phase-space. The total momentum and charge of the 
lattice vacuum are zero. The total negative-energy of the lattice vacuum is $-\pi/a_t$.
Upon this lattice vacuum, the Hamiltonian (\ref{efree}) is supposed to 
describe a positive energy excitation of massless left-handed fermion $\psi_L$ by 
the dispersion relation $E=-\tilde p, (\tilde p<0)$. 

The lattice vacuum, i.e., 
all occupied negative-energy and zero-energy states, must undergo quantum-field 
fluctuations at the scale of the lattice spacing (cutoff).
Due to these quantum-field fluctuations, the low-energy excitations of the 
massless particle (the left-mover $\psi_L$) and its antiparticle (a right-mover 
$\psi_L^c$) can be created from just below the zero-energy level $(E=0)$ of the 
lattice vacuum to just above, 
since they are massless and the energy-gap is zero. 
This is the phenomenon of pair-creations of massless particles, attributed to 
quantum-field fluctuations of the lattice vacuum. 
Quantum-field fluctuations are associated to energy-momentum fluctuations of
the lattice vacuum. 

At the scale of the lattice spacing, the energy fluctuation $|\Delta E|$ and momentum 
fluctuation $|\Delta \tilde p|$ of the lattice vacuum obey the Heisenberg uncertainty 
principle, 
\begin{equation} 
a_t|\Delta E|\simeq \pi,\hskip0.5cm a|\Delta \tilde p|\simeq \pi.
\label{uncer} 
\end{equation}
The energy fluctuation $|\Delta E|$ and momentum fluctuation $|\Delta \tilde p|$ of the 
lattice vacuum must be equal to the energy and momentum of particle and antiparticle 
created, since total energy and momentum of the lattice vacuum are conserved. 
This means that 
\begin{equation}
\Delta E=E_L^c-E_L \hskip0.2cm {\rm and}\hskip0.2cm 
\Delta \tilde p = \tilde p_L^c-\tilde p_L,
\label{fep}
\end{equation} 
where $E_L^c,\tilde p_L^c$ ($E_L,\tilde p_L$) are energy and momentum of the 
right-mover $\psi_L^c$(the left-mover $\psi_L$).
If the left-mover ($\psi_L$) is assumed at the energy-momentum state:
\begin{equation}
E_L=-\tilde p_L\simeq 0 \hskip0.2cm {\rm and}\hskip0.2cm 
\tilde p_L\simeq 0.
\label{cep}
\end{equation} 
Its antiparticle (the right-mover $\psi_L^c$) 
must be at the energy-momentum state:
\begin{equation}
E_L^c\simeq -\pi/a_t \hskip0.2cm {\rm and}\hskip0.2cm 
\tilde p_L^c\simeq -\pi/a,
\label{1cep}
\end{equation}
as resulted from eq.(\ref{uncer}).
However, this negative-energy state $E_L^c= - \pi/a_t$ is actually an empty state (hole) 
of the lattice vacuum. All fully filled negative-energy states with energy $E\in(0,-\pi/a_t)$ 
must fluctuate down to fill the empty states of lower negative-energy levels. 
As a consequence, that empty state (hole) at the negative-energy level $E_L^c\simeq -\pi/a_t$ 
moves upward to $E_L^c\sim 0$. This is just an extra low-energy excitation (doubler) of 
the antiparticle (right-mover $\psi_L^c$) at the energy-momentum state:
\begin{equation}
E_L^c\sim 0 \hskip0.2cm {\rm and}\hskip0.2cm 
\tilde p_L^c\simeq -\pi/a.
\label{1cepr}
\end{equation}
Note that the Hamiltonian (\ref{efree}) is invariant under 
the charge-conjugation. This shows how the vectorlike fermion-doubling phenomenon appears 
as the pair-creation of massless fermion and antifermion 
by quantum-field fluctuations of the lattice vacuum. When gauge fields 
perturbatively coupling to chiral fermions are turned on, gauge anomalies 
associated to the left-handed and 
right-handed fermions must appear and cancel each other.    

We turn to discuss the case of a massive Dirac fermion $\Psi=(\psi_L,\psi_R)$
with a ``soft'' mass 
``$m$'', which is much smaller than the energy fluctuation 
$|\Delta E|\sim \pi/a_t$ at the scale of the lattice spacing. We assume that 
this Dirac fermion $\Psi=(\psi_L,\psi_R)$ is at the 
energy-momentum state:
\begin{equation}
E_d=-m \hskip0.2cm {\rm and}\hskip0.2cm 
\tilde p_d\simeq 0.
\label{dep}
\end{equation} 
This indicates that 
the left(right)-handed fermion $\psi_L$($\psi_R$) is at 
the energy-momentum state:
\begin{equation}
E_L(E_R)=-m \hskip0.2cm {\rm and}\hskip0.2cm 
\tilde p_L(\tilde p_R)\simeq 0.
\label{1dep}
\end{equation} 
Its antiparticle, the right(left)-handed fermion 
$\psi_L^c$($\psi_R^c$), must be at the energy-momentum state:
\begin{equation}
E_L^c(E_R^c)\simeq -\pi/a_t \hskip0.2cm {\rm and}\hskip0.2cm 
\tilde p_L^c(\tilde p_R^c)\simeq -\pi/a,
\label{2dep}
\end{equation} 
where the fermion mass is negligible for $m\ll \pi/a_t$, as discussed 
in previous paragraph. However, this negative-energy state $E_L^c(E_R^c)\simeq - \pi/a_t$ 
(\ref{2dep}) is actually an empty 
state (hole) of the lattice vacuum. All fully filled negative-energy states with 
$E\in(-m,-\pi/a_t)$ 
must fluctuate down to fill the empty states of lower negative-energy levels. 
As a consequence, 
that empty state (hole) at the negative-energy level $E_L^c(E_R^c)\simeq -\pi/a_t$ moves upward 
to $E_L^c(E_R^c)\sim -m$. This is just an extra low-energy excitation of the 
antiparticles right(left)-handed fermion $\psi_L^c$($\psi_R^c$) at 
the energy-momentum state:
\begin{equation}
E_L^c(E_R^c)\sim -m \hskip0.2cm {\rm and}\hskip0.2cm 
\tilde p_L^c(\tilde p_R^c)\simeq -\pi/a.
\label{3dep}
\end{equation} 
This shows an extra mode (doubler) of the Dirac 
fermion $\Psi^c=(\psi_L^c,\psi_R^c)$ at the energy-momentum state $E_d\sim -m$ and 
$ \tilde p_d\simeq -\pi/a$.  

It is worthwhile to mention the unphysical case where the mathematical limit 
$a_t\rightarrow 0$ and $a\rightarrow 0$ are taken, i.e., the lattice regularization is 
removed. In this situation, the total number of negative-energy 
states in the energy-momentum space of the vacuum becomes {\it infinite}. 
As a result, the low-energy excitations (doubler) 
Eqs.(\ref{1cepr}) and (\ref{3dep}) appear at $\tilde p\rightarrow\infty$.

All these discussions can 
be generalized to the lattice vacuum in four dimensions. We summerize three important features 
playing key roles in these discussions of pair-creations around the zero-energy 
level of the lattice vacuum leading 
to the vectorlike doubling:
\begin{description}

\item [(1)] The regularized vacuum of a quantized fermion-field theory 
is an ensemble of negative-energy states fully occupied by virtual massless 
fermions and antifermions preserving symmetries; 

\item [(2)] the total number of all these 
negative-energy states of the regularized 
vacuum is finite;
 
\item [(3)] quantum-field fluctuations of the regularized vacuum at 
the scale of the regularization lead to the pair-creation, and the regularized 
vacuum quantum-mechanically fluctuating to the lowest energy state.

\end{description} 
These arguments are also applicable to {\it any} regularized 
vacuum whose total number of negative-energy states is appropriately regularized 
to be {\it finite}. This means that the number of negative-energy states 
with negative-energy lower than $-\pi/a_t$ (cutoff) 
are completely eliminated rather than exponentially suppressed.

It is worthwhile to compare and contrast our discussions with the proof of the 
``No-Go'' theorem\cite{nn81} and intuitive discussions on 1+1 dimensional 
lattice\cite{nn91}. The key points of exactly demonstrating ``No-Go'' theorem are 
(i) continuous fermion-spectrum $E(p)$ in an entire Brillouin zone, due to the 
locality of a free fermion theory; (ii) the compact topology of the Brillouin zone, 
due to the period boundary conditions of the fermion-spectrum $E(p)=E(p\pm\pi/a)$ 
on the lattice. Applying these two points to a free fermion theory on 1+1 dimensional
lattice, 
they give intuitive discussions: (i) such a doubling phenomenon is attributed to 
the fact that the numbers of down going states and up going states, crossing the 
zero-energy level of the vacuum, must be the same; 
(ii) an external electric field drives these down and up going states so that 
gauge anomalies are intrinsically related to the fermion doubling phenomenon. 
The property of the compact topology of the Brillouin zone is closely related to 
the crucial feature (2) discussed in this paper. Whereas other crucial features (1) 
and (3) discussed in this paper have relationships with both the compact topology of the 
Brillouin zone and the property of continuous fermion-spectrum.
It should be emphasized that this paper interprets the fermion doubling phenomenon from another 
physically intuitive view and reason, which are based on the feature of quantum-field fluctuations 
of the vacuum leading not only to excitations of fermion pairs, but also minimizing the energy of 
vacuum state. These are fundamental principles for the vacuum as the
ground state of quantum field theories.       
  
On the basis of these discussions on the relationship between the 
vectorlike fermion-doubling and 
massless fermion pair-creation phenomena on the lattice vacuum, we understand that
the vectorlike fermion-doubling on the lattice vacuum of a quantum field theory for free fermions
is inevitable, for the quantum-field fluctuating nature of the lattice vacuum. 
To evade this vectorlike 
fermion-doubling, we probably have to modify the energy-spectrum of the regularized vacuum.
Searching for a chiral-gauge-symmetric approach, that properly regularizes the 
standard model on the lattice, has been greatly challenging to particle physicists for the 
last two decades\cite{gw}-\cite{xue99}.
One of possible ways is to introduce the 
strong interactions between fermions. These strong interactions change the energy-spectrum 
of the lattice vacuum in a such way that a chiral gauge invariant mass-gap at the 
order of $\pi/a_t$ is 
created only for extra fermion doublers, while desired chiral fermions are kept as massless 
excitations of $E=\pm\tilde p$.


\end{document}